\pdfoutput=1
\documentclass[reqno]{amsart}
\usepackage{ifpdf}
\usepackage{graphicx}
\usepackage{cite}
\usepackage{amsmath}
\usepackage{amssymb}
\usepackage{epstopdf}
\usepackage{pdfpages}
\usepackage{float}
\usepackage{color}

\usepackage[font=footnotesize]{caption}
\usepackage{subcaption}
\usepackage{multicol}
\usepackage[margin=1.5in]{geometry}
\begin{document}

\begin{center}
    \textbf{ \Huge Torsional oscillations of a sphere in a Stokes flow} \\[0.5 cm]
		F.\ Box$^{1,3}$, A.\ B.\ Thompson$^{2}$ and T.\ Mullin$^{1}$\\ 
		\textit{$^{1,2}$Manchester Centre for Nonlinear Dynamics, $^{1}$School of Physics and Astronomy and $^{2}$School of Mathematics, University of Manchester, Oxford Road, Manchester, M13 9PL, UK\\
$^{3}$Institute for Theoretical Geophysics, Department of Applied Mathematics and Theoretical Physics, University of Cambridge, Wilberforce Rd, Cambridge,\\ CB3 0WA, UK }\\

\end{center}

\noindent The results of an experimental investigation of a sphere performing torsional oscillations in a Stokes flow are presented. A novel experimental set up was developed which enabled the motion of the sphere to be remotely controlled through application of an oscillatory magnetic field. The response of the sphere to the applied field was characterised in terms of the viscous, magnetic and gravitational torques acting on the sphere. A mathematical model of the system was developed and good agreement was found between experimental and theoretical results. The flow resulting from the motion of the sphere was measured and the fluid velocity was found to have an inverse square dependence on radial distance from the sphere. Agreement between measurements and the analytical solution for the fluid velocity indicates that the flow may be considered Stokesian.


\section{Introduction}
\indent The torsional oscillations of a sphere in a viscous fluid is a classical problem in fluid mechanics which has received significant theoretical study but relatively few experimental investigations since it was first considered in 1860 \cite{Helmholtz1860}. Moreover, all the experimental work undertaken was performed on torsion pendulums - spheres supported in the fluid by a mechanically driven rod or torsion fibre - and did not consider the viscous effects introduced by the mechanical support \cite{Hollerbach2002, Folse1974, Folse1994, Benson1956}. The effects of an inelastic tether on the motion of a particle in an oscillating flow have been shown to be important at low frequencies of oscillation \cite{Lim2014}, while the viscous drag introduced by a string fastened to a sphere moving in a fluid has also been reported \cite{Chen1994}. 

We have therefore developed a remote control approach to investigating the textbook \cite{Landau1987} problem of a free, torsionally oscillating sphere at low Reynolds number ($Re = \omega a^{2}/\nu$, where $\omega$ is the angular frequency, $a$ is the radius of the sphere and $\nu$ is the kinematic viscosity) and low Strouhal number ($St = a\omega/9V$, where $V$ is the magnitude of the fluid velocity).

In the classic monograph \textit{Hydrodynamics}, Lamb \cite{Lamb1932} derived the analytic solution for a steadily rotating sphere, at $Re=0$, and for a sphere performing small-amplitude torsional oscillations, at low but non-zero $Re$. In both cases, the sphere was assumed to be free and either in an infinite domain or rotating inside a larger concentric sphere, and the fluid was shown to move around the oscillating sphere in concentric shells, the centres of which lie on the axis of rotation. Similarly, for small-amplitude oscillations and negligible secondary fluid motion, the flow resulting from the decaying oscillations of a torsion pendulum was determined to be circumferential in annuli around the axis of rotation \cite{Buchanan1891, Kestin1954}. Tekasul \textit{et al}. \cite{Tekasul1998} numerically solved for the torque on a torsionally oscillating sphere in an unbound medium using a Green's function approach and found a less than $0.1\%$ difference with the analytic solution of Lamb. Kanwal \cite{Kanwal1955} investigated the oscillatory rotation of rigid axi-symmetric bodies about an axis of symmetry in a viscous, incompressible fluid using Stoke's stream function. Lei \textit{et al}. \cite{Lei2006} calculated the viscous torque on a spherical particle under arbitrary rotation and a particle rotating in a velocity field rather than a still fluid. By considering a characteristic dimension of the body, Zhang and Stone \cite{Zhang1998} extended the analysis to the oscillatory translations and rotations of nearly spherical particles.  

Considering $Re>1$, Carrier and Di Prima \cite{Carrier1956} used perturbation techniques to solve the nonlinear Navier-Stokes equations which result when the radial and axial flows are not neglected and uncovered a secondary circulatory motion in planes containing the axis of rotation: fluid recedes from the sphere along the equatorial plane and flows in at the poles. This inertial effect results from a centrifugal force which is greatest at the equator. For increasing $Re$, the streaming effect increases and a radial jet results from hemispherically-symmetric circulatory flows colliding at the equator. The radial jet is ejected along the equatorial plane and, for large-amplitude oscillations, generates vortex pairs which break down into turbulence \cite{Hollerbach2002}.

The experimental work undertaken \cite{Folse1974, Folse1994, Benson1956} focused on the damping of a torsion pendulum. In particular, a sphere was suspended in the viscous fluid, deflected from the equilibrium position and subjected to the restoring couple of the supporting fibre and the surrounding fluid such that it approached the equilibrium position through decaying oscillations. The experiments of Folse \cite{Folse1974}, in particular, led to the identification of an algebraic error in the original calculation of the correction term to the torque which, for $Re>1$, results from the development of circulatory streaming motion \cite{DiPrima1976}. Hollerbach \textit{et al}. \cite{Hollerbach2002}, in contrast, used a sphere supported from above and below by a mechanical rod, and driven to oscillate with constant time-period, to examine the radial jet which results from colliding circulatory flows. Flow visualisation of the laminar to turbulent transition of the radial jet provided an explanation of the abrupt variation in the damping rate of the torsion pendulum found by Benson and Hollis Hallett \cite{Benson1956} in experiments performed in liquid helium. Although tethering conflicts with the theoretical assumption of a free sphere, the effect of the motion of the supporting fibre or rod was not considered in the aforementioned experimental investigations. The authors are not aware of any experimental work conducted on a free sphere performing torsional oscillations, a fact which results from the inherent difficulty present in driving the motion of a body without mechanical contact.

Here, we present a novel experimental system which was developed to investigate the fluid motion induced by spheres which were forced to oscillate without mechanical contact in an experimental realisation of a Stokes flow. Near neutrally-buoyant spheres, with a magnetic dipole axis, were submerged in a very viscous fluid. Application of a steady magnetic field introduced a magnetic torque which acted to align the magnetic dipole of the sphere with the applied field causing the sphere to rotate. Through application of an alternating magnetic field, the sphere was made to perform torsional oscillations with an amplitude and a frequency determined by the magnitude and frequency of the applied field. 

The fluid flow generated by a magnetic sphere in a viscous fluid subject to an applied magnetic field has a variety of industrial and biomedical applications. Although the following applications utilised micro-particles composed entirely of magnetic material, rather than macro-particles containing magnets, comparisons can be made between the non-Brownian dynamics of the particles and the sphere considered in this work and, as such, potential applications, as outlined below, are envisaged. 

The rotational fluid flow generated by a magnetic micro-sphere driven by an external rotating magnetic field can propel passive micro-objects, and therefore be used for precise non-contact manipulation and long-range transportation of micro-objects. Multiple micro-manipulators in parallel create reconfigurable, virtual micro-fluidic channels for concurrent, non-contact transportation of multiple micro-objects \cite{Ye2012}. Similarly, the hydrodynamic flow generated by paramagnetic colloidal particles subject to an external, rotating magnetic field has been used as a micro-stirrer to mix colloidal suspensions \cite{Tierno2007}. The rotation of a magnetic micro-sphere subject to an external rotating magnetic field has been studied and the asynchronous rotation of the micro-sphere with the driving field, which occurs above a critical frequency of magnetic field, can be used to detect and monitor bacterial growth \cite{McNaughton2007} \cite{McNaughton2009}.

As well as being used to mix fluids and manipulate micro-objects, the hydrodynamic forces generated by the motion of a magnetic sphere subject to an applied magnetic field can also be used to measure rheological properties of the fluid \cite{Sakai2010}. This includes non-invasive \textit{in vivo} viscosity measurements \cite{Besseris1999}. Furthermore, with the advent of nanotechnology, magnetic nano-particles (MNPs) and colloidal suspensions of magnetic particles have been utilised in a wide range of biomedical applications \cite{Yang2012} \cite{Pankhurst2003}. The controlled manipulation, through the application of external magnetic fields, and the size compatibility of MNPs with biological cells mean they are an important tool in both \textit{in vitro} and \textit{in vivo} applications. \textit{In vitro} applications include the detection, separation and monitoring of biological species, and blood purification. \textit{In vivo} applications include Magnetic Resonance Imaging (MRI), site-specific drug delivery and treatment of hyperthermia \cite{Medeiros2011}.

The experimental set up, which was designed to remotely control the oscillatory motion of a sphere in a viscous fluid, is detailed in \S\ref{sec:exp}. The mathematical model of the sphere response to the applied field is described in \S\ref{sec:theory}. Characterisation of the dynamic response of the sphere to magnetic actuation and measurements of the flow induced by the motion of the sphere are detailed in \S\ref{sec:results}\ref{subsec:resultssphere} and \S\ref{sec:results}\ref{subsec:resultsfluid}, respectively.

\section{Experimental Set Up}
\label{sec:exp}

\begin{figure}
	\centering
		\includegraphics[width=0.75\textwidth]{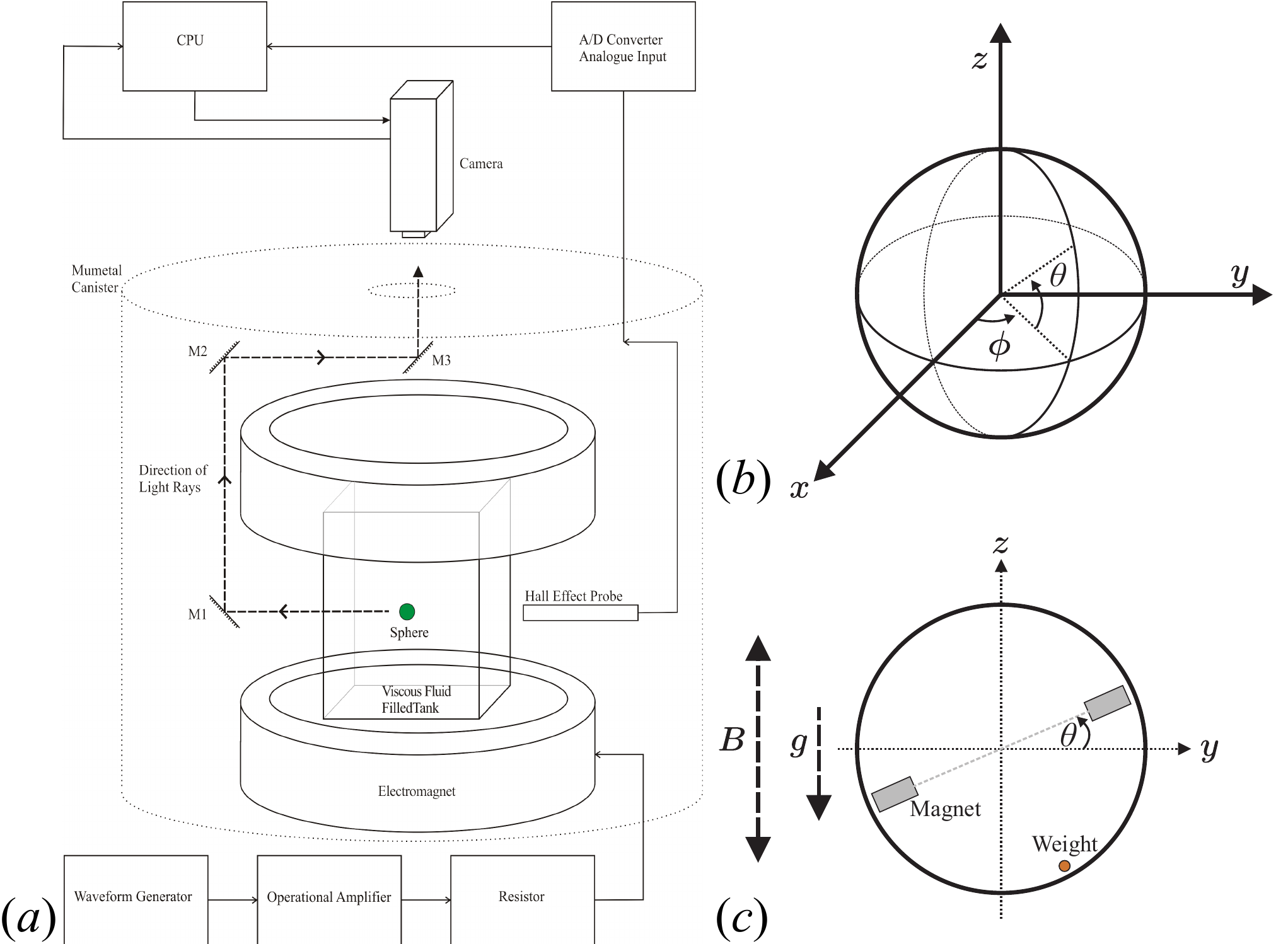}
			\caption{(a) Schematic diagram of the experimental apparatus. Near neutrally-buoyant spheres in a viscous fluid filled tank were imaged using an optical arrangement of mirrors (M1, M2 and M3) and a computer-controlled camera. A magnetic field was applied using Helmholtz coils and measured by a Hall effect probe connected to a PC via an 16-bit analogue-to-digital signal converter. The current supplied to the Helmholtz coils was modulated using a waveform generator, amplified by an operational amplifier and passed through a resistor which acted as a resistive load. The experimental system was contained within a Mumetal canister to reduce the effects of extraneous magnetic fields. (b) Schematic diagram of the coordinate system, and (c) a projection of the coordinate system onto the observed plane. The projection of the coordinate system also includes indications of the direction of the applied magnetic field, $B$, which interacts with the magnets embedded in the sphere, and the direction of gravity, $g$, which induces a torque on the non-uniform mass distribution within the sphere.  }
		\label{fig:Sch1}
\end{figure}

A schematic diagram of the experimental set up used to characterise the dynamic response of the sphere to an applied magnetic field is shown in Figure \ref{fig:Sch1}(a). Experiments were conducted on a polypropylene sphere of diameter $2a =15.86\pm0.01$ mm (Dejay Distribution Ltd., UK). Two neodymium, permanent magnets were inserted into machined holes in the sphere and small amounts of glue were used to hold them in place. The two magnets were carefully embedded in the sphere such that they were flush with the surface and diametrically opposite ($\pm1^{\circ}$) along an axis through the centre of the sphere. The permanent magnets were cylindrical, of length $3.00\pm0.01$ mm and diameter $2.48\pm0.01$ mm, occupying less than $2$\% of the sphere volume. The adjacent poles of the magnets were placed opposite one another such that the two magnets acted as a single dipole, the length of which was equal to the sphere diameter. The permanent magnets had a remanence of $1.19\pm0.05$ T and a coercivity $\geq868$ kA\,m$^{-1}$ \cite{Yunsheng}. These values indicate that the strength of the magnets is large for their size compared to standard ferromagnetic materials such as iron. The magnetic dipole moment of an individual magnet, $m = 0.0140\pm0.0006$ A\,m$^{2}$, was effectively independent of the applied magnetic field for the range of values used \cite{Jackson1998}.

After the inclusion of the magnets, additional weights were embedded in the sphere in order to obtain near neutral buoyancy. The zero-field orientation of the magnetic dipole of the sphere was controlled by careful positioning of the embedded weights such that the magnetic dipole axis of the sphere was orientated approximately orthogonal to the applied field direction, at an angle of $3.22\pm0.77^{\circ}$ from the horizontal position. The resulting density of the sphere was $\rho_{s} = 978.53\pm3.8$x$10^{-3}$ kg\,m$^{-3}$ but the sphere had a non-uniform mass distribution that introduced a gravitational torque.

The centre of mass was not exactly at the centre of the sphere, but instead was displaced a distance $\lambda a$ from the centre, and at an angle $\theta_0$ in the projected plane depicted in Figure \ref{fig:Sch1}(c). The exact calculation of $\lambda$ (in the range 0 to 1) and $\theta_0$ would require complete knowledge of the density distribution within the sphere. Instead we use an empirical calibration based on the rate at which the sphere returns to $\theta=\theta_0$ if displaced from equilibrium. This calibration method is discussed further in \S\ref{sec:theory}\ref{subsec:gravtime}. We find that $\theta_0=3.09^\circ$ and $\lambda =0.016$, so that the centre of mass is displaced only very slightly from the centre of the sphere. However, the resulting gravitational moment due to the offset centre of mass can still affect the dynamics of the ball in response to the imposed magnetic field. The time-scale associated with the gravitational torque acting on the sphere was measured, using the method outlined in \S\ref{sec:theory}\ref{subsec:gravtime}, and found to be $T_{0} = 4.67\pm0.03$ s.

The sphere was submerged in a viscous liquid inside a rectangular tank, made of $5$ mm thick perspex with internal dimensions of width $125$ mm, length $115$ mm and height $200$ mm. The centre of the sphere $\approx 8a$ from the side walls of the tank, such that the correction to the torque resulting from the boundaries was at most $0.2\%$. Experiments were performed in a temperature-controlled laboratory and the temperature inside the system was measured to be $T=19.89\pm0.30$ $^{\circ}$C. The working fluid was silicone oil (Basildon Chemical Company Limited, UK) with a measured viscosity of $\nu = 924.14\pm5.29$ mm$^{2}$\,s$^{-1}$ and a measured density of $\rho_{f} = 975\pm1$ kg\,m$^{-3}$. The fluid-filled tank was positioned on a platform of adjustable height in the centre of Helmholtz coils. The height and thickness of each Helmholtz coil was $140$ mm and $35$ mm respectively, and the gap between the two coils was $55$ mm. The electromagnet inductance was $50$ mH with a DC resistance of $0.36\pm0.04$ $\Omega$. A spatially uniform, alternating magnetic field was applied in order to study the angular response of the sphere from the zero-field position. Hall effect probes were used to measure the applied magnetic field, the amplitude, $B$, and frequency of which varied from $0$ to $\sim2.5$ mT and from $0.01-4$ Hz, respectively. The amplitude and frequency of the applied field were chosen such that the Reynolds number of the flow was in the range $0.01 < Re < 0.1$ and the Strouhal frequency $St < 0.2$. 

Shielding from the Earth's magnetic field was achieved by placing the experimental system inside a Mumetal canister. Mumetal is a nickel-iron alloy with a magnetic permeability over 100 times greater than steel and shields by providing a path of low reluctance, and thus entraining magnetic flux. The Mumetal shield consisted of a $1.60\pm0.02$ mm thick cylindrical container, diameter and length of $510$ mm and $520$ mm respectively, with a lid and a base plate. Both the lid and base plate contained a centrally-located hole of $70$ mm diameter through which cables could be passed to connect the experimental apparatus to control boxes and a PC, and permit observations of the motion of the spheres. The magnetic field inside the Mumetal shield was consistently found to be less than $5$ $\mu$T, more than one order of magnitude less than the background field in the laboratory that was comparable to the geomagnetic field, which varies from $25$ to $65$ $\mu$T \cite{MagneticIntensity}.

The sphere was illuminated using two $400$ mm strips of $36$ Light Emitting Diodes (LEDs). The strips were attached to the inside surfaces of the upper and lower coils of the electromagnet to illuminate the region of interest from above and below. This set-up produced uniform illumination across the observable surface of the sphere. Furthermore, prior to each experiment, the rotational plane of the sphere was carefully orientated to align with the observational plane using a laser-sheet as a guide. The motion of the sphere was recorded using a Genie camera (HM-1400, Teledyne DALSA, Canada), with a spatial resolution of $1400$x$1024$ pixels, which was positioned above the central hole of the lid of the Mumetal canister, as shown in Figure \ref{fig:Sch1}(a). The light transmitted through the tank was directed to the camera in a periscope-like manner via three mirrors (M1, M2 and M3), of width $76$ mm and height $102$ mm, positioned in between the electromagnet and the Mumetal canister. Mirror M1 was placed at $45^{\circ}$ with respect to the side of the tank through which the sphere was observed. Mirrors M2 and M3 were adjusted in order to create an image at the camera which was centered on the CCD sensor.  Images of the sphere motion were typically acquired at a frame rate 100 times greater than the frequency of the applied field, except for frequencies $>0.6$ Hz for which the images were acquired at the maximum frame rate of $64$ Hz. The image resolution was $0.02$ mm/pixel and the maximum recorded drift velocity of the sphere was $\approx 0.1$ mm/min.

Analysis of the dynamic behaviour of the sphere involved tracking its motion. Schematic diagrams of the coordinate system and the observed plane are shown in Figures \ref{fig:Sch1}(b) and \ref{fig:Sch1}(c), respectively. The sphere used was coloured black with white markings, using permanent marker pens. Images of the sphere were processed using MATLAB software (R2011a, MathWorks Inc., USA). Tracking the motion of sphere throughout image sequences enabled the construction of a time-series of the dynamic response of the sphere to the applied magnetic field.

\begin{figure}
	\centering
		\includegraphics[width=0.75\textwidth]{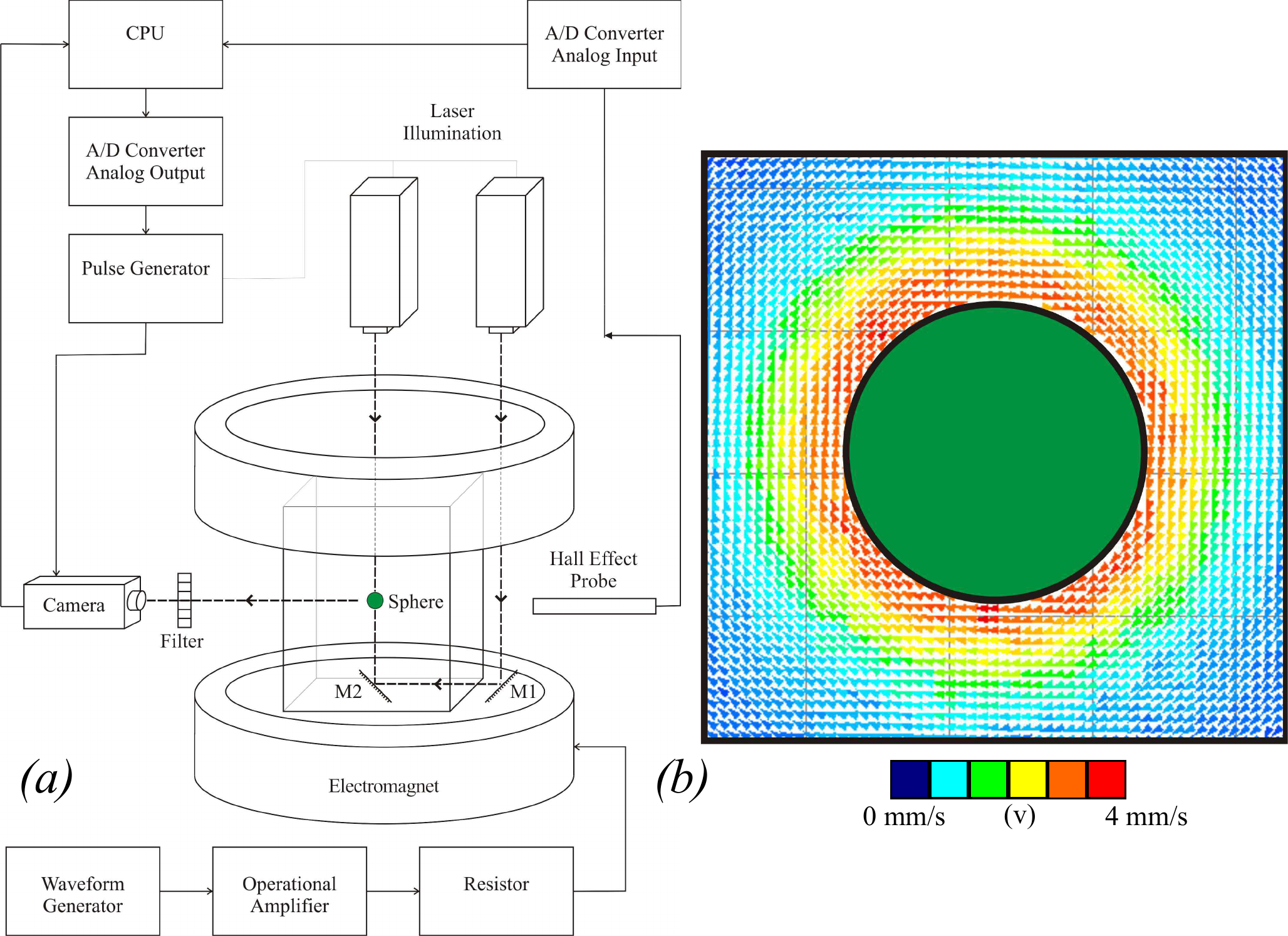}	
			\caption{(a) Schematic diagram of the experimental apparatus used for flow visualisation. Illumination of the rotational plane of the sphere was provided by lasers and an optical arrangements of mirrors (M1 and M2). A high-speed camera was positioned orthogonal to the illuminated plane and a low-pass filter was positioned between the tank and the camera to remove noise from the signal. (b) A typical velocity vector field depicting the flow generated by a spherical particle rotating, in the clockwise direction, in a viscous fluid. The instantaneous velocity field was measured using Particle Image Velocimetry at $3\pi/10$ in the oscillation cycle. The green circle in the centre of the image denotes the sphere. The vectors are represented by coloured arrows, the magnitude of which ranges from $0$ mm\,s$^{-1}$ (dark blue) to $4$ mm\,s$^{-1}$ (dark red).}
		\label{fig:Sch2PIV}
\end{figure}

A schematic diagram of the experimental apparatus used for flow visualisation is shown in Figure \ref{fig:Sch2PIV}(a). Neutrally-buoyant, spherical micro-particles (Fluostar particles, EBM Corporation, Japan) of $13.9$ $\mu$m mean diameter were suspended in the viscous fluid and had a Rhodamine B coating which fluoresced under green-light laser illumination of wavelength $532$ nm. A cross-section of the tank, corresponding to the rotational plane of the driven sphere, was illuminated using two green-light lasers. A continuous $50$ mW laser sheet illuminated the cross-section from above, whilst a Nd:YAG pulsed laser illuminated the cross-section from below after conversion from a laser dot to a sheet by passing through a cylindrical lens. The Nd:YAG laser was positioned next to the $50$mW laser, above the tank, and illuminated the cross-section from below after reflecting off two $45$ degree mirrors, one positioned within the tank and one outside. A semi-cylindrical lens was used to reduce the width of the laser sheet to $\approx{1}$ mm to focus the illuminated plane on the centreline of the sphere. A low-pass filter was positioned between the tank and the camera to reduce background noise in the detected signal. A high-speed camera (pco.1200 hs, PCO AG, Germany) with a spatial resolution of $1280$x$1024$ pixels was used to image the region of interest. The camera was positioned orthogonal to the illuminated plane, synchronised with the Nd:YAG pulsed laser using a pulse generator (BNC Model 500, Oxford Lasers Ltd., UK) and imaged at a rate of $15$ Hz, the maximum pulse-rate of the Nd:YAG laser, with an exposure of between $10$ and $20$ ms. Pairs of consecutive images were then subject to sub-pixel Particle Image Velocimetry. A consequence of the maximum pulse rate of the laser was that flow visualisation measurements were conducted for a frequency of applied magnetic field of $0.15$ Hz.

A typical, instantaneous velocity vector field, obtained at $3\pi/10$ in the oscillation cycle of the sphere, is shown in Figure \ref{fig:Sch2PIV}(b). The sphere was rotating in the clockwise direction which induced rotation of the surrounding fluid in the same direction. The velocity vectors are represented by arrows, the colour of which denotes the magnitude of the velocity which ranges from $0$ mm\,s$^{-1}$ to $4$ mm\,s$^{-1}$. The velocity vector field shows that the velocity of the fluid is greatest close to the surface of the sphere and decreases with radial distance from the sphere surface.

\section{Theory}
\label{sec:theory}

The dynamic response of a sphere, containing a magnetic dipole, to an applied magnetic field is considered theoretically in this section. A schematic of the coordinate system used in the development of the model is shown in Figure \ref{fig:Sch1}(b). In the experiments, the orientation of the rotational plane of the sphere was carefully set to be orthogonal to the observer such that $\phi=0$, where $\phi$ is the angle between the magnetic-dipole axis of the sphere and the line-of-sight of the observer - the $x$-axis. A projection of the coordinate system in the observation plane is shown in Figure \ref{fig:Sch1}(c). Also indicated in Figure \ref{fig:Sch1}(c) are the direction of gravity, $g$, the direction of the applied magnetic field, $B$, and the position of the effective magnetic dipole within the sphere. The orientation of the magnetic dipole of the sphere with zero applied field is approximately horizontal in the eyes of the observer, $\theta\sim0$ degrees where $\theta$ is the angle of the magnetic-dipole axis of the sphere from the $y$-axis. 

Under the assumptions that $Re=0$ and that the sphere is rotating in unbounded fluid, the equation of motion for the dynamic response of a sphere subject to an applied magnetic field can be deduced from a combination of the viscous, gravitational and magnetic torques acting on the sphere:
	\begin{equation}
		8\pi\mu a^{3}\frac{d\theta}{dt} = -\epsilon \sin\theta + Bm\sin\omega t\cos\theta,
			\label{eq:torquebalance}
	\end{equation}	
where $\mu$ is the dynamic viscosity of the fluid, $a$ is the radius of the sphere, $B$ is the magnetic field strength, $m$ is the magnetic moment of the magnetic-dipole of the sphere, $\omega$ is the angular frequency, $\theta$ denotes the orientation of the magnetic axis of the sphere in the observational plane and $\epsilon$ is the gravitational torque which acts to return the sphere towards the zero applied field orientation at $\theta\sim0$. Based on centre of mass arguments, $\epsilon$ can be considered to be equal to $M g \lambda a$, where $M = 4\pi a^3 \rho_{s}/3$.The viscous torque is a result of the no-slip condition which applies on the surface of the sphere and provides hydrodynamic resistance to the motion of the sphere. The gravitational torque arises from gravity acting on the non-uniform mass distribution within the sphere and acts to return the magnetic-dipole of the sphere to the zero-field position.  The magnetic torque results from the interaction of the magnetic dipole of the sphere and the applied field and acts to align the magnetic-dipole of the sphere with the applied field. 

A non-dimensional time, based on the angular frequency $\omega$ of the applied oscillatory field, can be defined as $\hat{t} = \omega t$. Equation \ref{eq:torquebalance} then becomes
	\begin{equation}
		\frac{d\theta}{d\hat{t}} = -\frac{\epsilon}{8\pi\mu a^{3}\omega}\sin\theta + \frac{Bm}{8\pi\mu a^{3}\omega}\sin\hat{t}\cos\theta.
	\end{equation}
Dimensionless parameters which quantify the ratio of the gravitational torque to the viscous torque, $\epsilon$, and the viscous torque to the magnetic torque, $Ma$, acting on the sphere are thus defined by:
\begin{equation}
		\hat{\epsilon} = \frac{1}{\omega T_{0}},     \quad	\frac{1}{Ma} = \frac{Bm}{8\pi\mu  a^{3}\omega}, 
	\end{equation}
respectively, where $T_{0} = 8\pi\mu a^{3}/\epsilon$ is the gravitational time-scale, discussed in further detail in \S\ref{sec:theory}\ref{subsec:gravtime}, and $Ma$ is the Mason number \cite{Man2013}. This Ordinary Differential Equation (ODE) for $\theta$ can thus be written as 
	\begin{equation}
		\frac{d\theta}{d\hat{t}} = -\hat{\epsilon}\sin\theta +\frac{1}{Ma}\sin\hat{t}\cos\theta.
			\label{eq:ODEtheta}
	\end{equation}
		
We can solve equation \ref{eq:ODEtheta} numerically by direct integration; this requires a single initial condition, for example the value of $\theta$ when $t=0$.
If $\hat{\epsilon}=0$, then we can separate equation \ref{eq:ODEtheta} and integrate to find $\theta(t)$ analytically. All solutions for $\theta(\hat{t})$ are periodic with period $2\pi$, and form a two parameter family characterised by the mean value of $\theta$ and the Mason number $Ma$. When $\hat{\epsilon}>0$, the solution $\theta(t)$ is not usually strictly periodic, but over several cycles the solution converges towards a limit cycle for which $\bar{\theta}=0$. In the limit $\hat{\epsilon}\rightarrow 0$, we retain the condition that $\bar{\theta}=0$.
The solution for the equations with $\hat{\epsilon}=0$ and $\bar{\theta}$ satisfies $\theta_{min}=-\theta_{max}$ and 

	\begin{equation}
		Ma^{-1} = \ln(\sec(\Delta\theta/2)+\tan(\Delta\theta/2)).
			\label{eq:analyticsol}
	\end{equation}.
	
When $\hat{\epsilon}>0$, we cannot solve equation \ref{eq:ODEtheta} analytically, but we can determine the periodic limit cycles of $\theta(\hat{t})$ by solving a boundary value problem in Matlab (R2011a, Mathworks, USA) with periodic boundary conditions; the resulting limit cycles always have mean zero, but do not satisfy equation \ref{eq:analyticsol}.



\subsection{Gravitational Time-scale}
\label{subsec:gravtime}

Individual spheres have unique, non-uniform mass distributions, an inevitable result of the embedding of magnets and weights. This non-uniform mass distribution supplies a net gravitational torque $\epsilon$ which acts to return the sphere towards a preferred orientation $\theta=\theta_0$. The embedded weights were positioned so that $\theta_0$ is close to zero, and so the magnetic dipole is close to horizontal when zero magnetic field is applied.

In order to characterise the dynamic response of a sphere to the applied field a precise measure of the gravitational torque was required. A simple test involved measuring the rate at which
the sphere returns to the zero-field position when released from a non-zero angle in the absence of magnetic forcing. With $B=0$, equation \ref{eq:torquebalance} becomes

\begin{equation}
	\label{full-decay-equation}
	\frac{d\theta}{dt}= - \frac{\epsilon}{8\pi \mu a^3} \sin{\theta}
\end{equation}
For small displacements, $\sin\theta \sim \theta$, and so $\theta \propto \exp(-t/T_0)$, with time constant

\begin{equation}
	\label{linearised-decay-equation}
	T_0 = \frac{8\pi \mu a^3}{\epsilon}.
\end{equation}

Estimates of the time-scale associated with the gravitational torque were obtained by measuring the decay of the angular position of a sphere from an offset position towards the zero-field state upon sudden removal of the applied magnetic field. A large ($\sim 2.2$ mT) steady magnetic field was applied initially in order to attain alignment of the magnetic dipole with the applied field. Sudden removal of the applied field led to the rotation of the sphere back to the zero-field position under the influence of gravity. A time-series of the angular position of the sphere was obtained from an initial angle of $30^\circ$ to the zero-field orientation, as shown in Figure \ref{fig:GdecayPhaseTimeSeries}(a). A least-squares fit of the form $\theta = A \exp(-t/T)+B$ yielded $A=28.21^\circ$, $B=3.09^\circ$ and $T=4.67$ s. The initial angle of $30^\circ$ is sufficiently small that \eqref{full-decay-equation} and \eqref{linearised-decay-equation} yield almost indistinguishable predictions for $\theta(t)$, and so the small angle approximations that led to exponential decay are valid. The accuracy of the measurement of the time-scale $T_0$ associated with the gravitational torque was improved by calculating the mean time-scale from 20 measurements. The time-scale associated with the discharging of voltage in the electromagnets was orders of magnitude faster than $T_0$ and was therefore neglected.

The importance of the gravitational torque in our model is expressed through the dimensionless parameter $\hat{\epsilon}=1/(\omega T_0)$. In our experiments, the frequency of the magnetic field ranged from $0.01$ to $4$ Hz, yielding $\hat{\epsilon}$ the range 0.009 to 3.5. Thus gravitational torque is typically not negligible in our experiments.

This measurement of $T_0$ not only allows us to calculate $\hat{\epsilon}$ but also supplies information regarding the mass distribution within the ball. We note that
\begin{equation}
\epsilon = \frac{8\pi \mu a^3}{T_0} = \frac{4\pi a^4 \rho_{s} g \lambda}{3}
\end{equation}
where $\lambda a$ displacement of the centre of mass of the sphere. With $T_0=4.67$ s, we find that $\lambda =0.0152$, and so the centre of mass is only slightly offset from the centre of the sphere.

\section{Results}
\label{sec:results}

\subsection{Magnetic actuation and response of the sphere}
\label{subsec:resultssphere}

\begin{figure}
		\centering
		\includegraphics[width=0.85\textwidth]{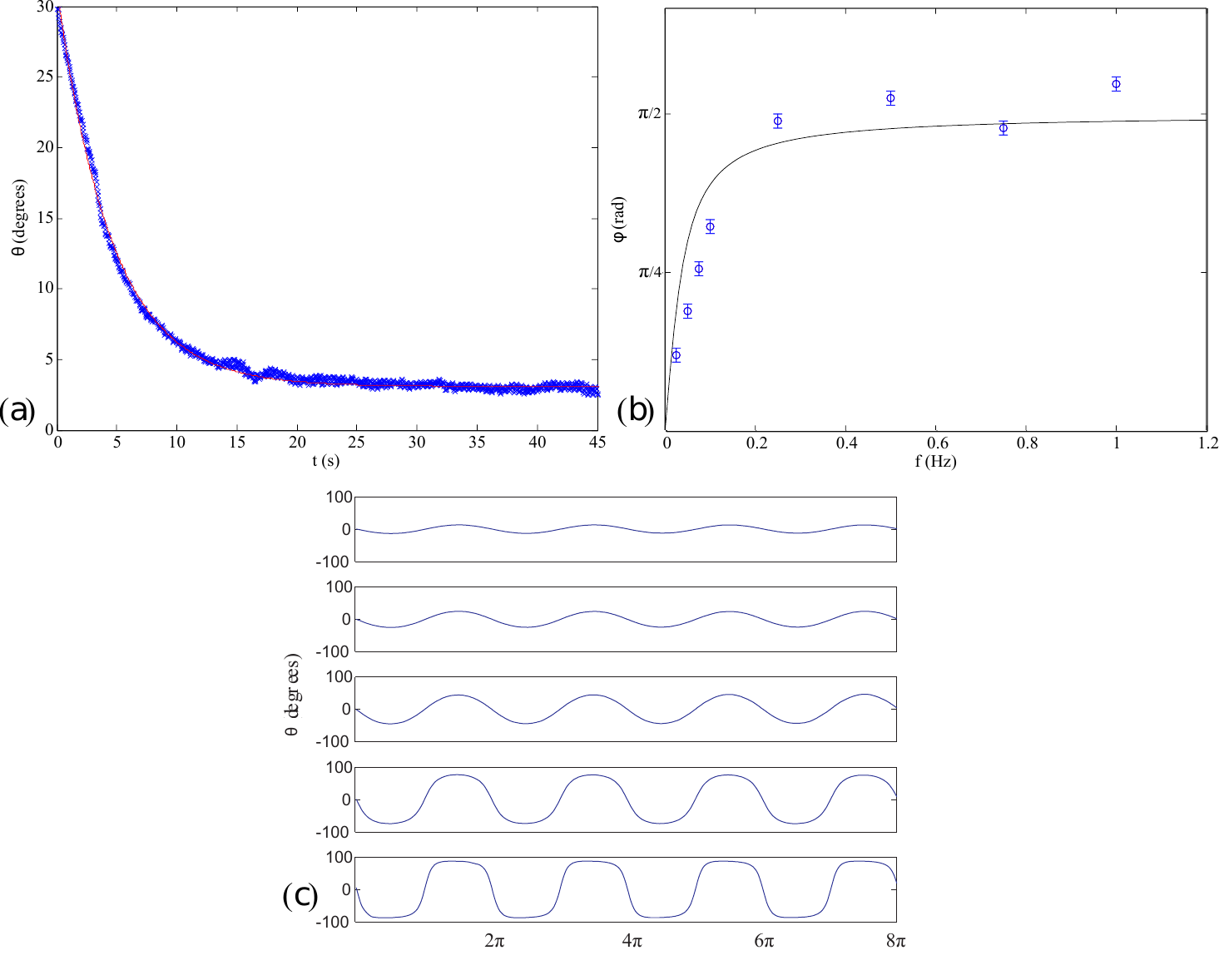}
			\caption{(a) A time-series of the angular position of the magnetic-dipole axis of the sphere measured as a function of time as the sphere rotates back to the zero-field orientation under the sole influence of gravity. A least-squares fit of the data to $\theta = A \exp(-t/T)+B$ yielded $A=28.21^\circ$, $B=3.09^\circ$ and $T_{0}=4.67$ s. (b) The phase delay, $\varphi$, between the applied magnetic field and the resulting motion of the sphere given by $\varphi=\arctan(2\pi T_0 f)$, depicted by the solid, black line, and measured experimentally (\textcolor{blue}{o}) as a function of the frequency of applied field, for values of $Ma$ and $\hat{\epsilon}$ such that $\theta \approx 16^{\circ}$. (c) Experimental time-series of the angular displacement of the driven, magnetic sphere over four periods of oscillation, for $\hat{\epsilon}=0.22$ and $Ma = 4.85$, $2.43$, $1.21$, $0.48$ and $0.20$ (top-to-bottom). The response of the sphere deviates increasingly from the form of the sinusoidal drive with decreasing $Ma$.}
		\label{fig:GdecayPhaseTimeSeries}
\end{figure}

The phase delay, $\varphi$, between the applied magnetic field and the response of the sphere is given by $\varphi=\arctan(2\pi T_0 f)$. This result is obtained by linearising equation \ref{eq:ODEtheta} for small $\theta$ and seeking a solution with $\theta = A \sin(t+\phi)$. The deduced phase delay and experimentally measured values of $\varphi$ are shown as a function of frequency of applied field, $f$, in Figure \ref{fig:GdecayPhaseTimeSeries}(b). The phase delay increases from $<\pi/10$ at a frequency of $0.025$ Hz and saturates at $\approx \pi/2$ for frequencies above $0.25$ Hz. Typically, measurement of the dynamic response of the sphere to the applied field were obtained at a frequency of $0.5$ Hz for which the phase delay was $\approx\pi/2$.

Experimental time-series of the angle of the sphere, $\theta$, are shown in Figure \ref{fig:GdecayPhaseTimeSeries}(c), for $\hat{\epsilon}=0.22$ and  $Ma = 4.85$, $2.43$, $1.21$, $0.48$ and $0.20$ (top-to-bottom), which corresponds to applied magnetic fields of amplitude $0.112\pm0.015$, $0.224\pm0.015$, $0.448\pm0.015$, $1.122\pm0.015$ and $2.688\pm0.015$ mT, respectively. Four periods of oscillation are shown in each case. 

When $Ma$ is large, the induced magnetic torque is small compared with the viscous torque resisting motion, and so the sphere performs only small-amplitude torsional oscillations about the zero-field orientation. The oscillations were sinusoidal and matched the waveform of the applied field such that the response of the sphere was linearly proportional to the magnetic forcing. For decreasing $Ma$, the amplitude of oscillation increased nonlinearly. Furthermore, the response of the sphere deviated increasingly from the sinusoidal form of the applied field as alignment of the magnetic axis with the applied field in both direction was approached. A sinusoidal function was fitted to the experimental data using a least-squares method and, for oscillations of amplitude less $19^{\circ}$, the standard deviation between the fitted function and the experimental data was less than $0.1\%$. Moreover, for oscillations of amplitude less than $45^{\circ}$, which resulted for $Ma\approx1.2$, the standard deviation between the data and the fitted sinusoidal function was less than $1\%$. 

For small Mason number, $Ma\lesssim1$, the magnetic torque acting on the sphere dominated over the viscous resistance and the sphere performed large-amplitude oscillations which deviated significantly from the sinusoidal form of the applied field. Small $Ma$ resulted in a nonlinear response of the sphere as the magnetic dipole quickly attained approximate alignment with the applied field, saturating the angular response at $\pm90^{\circ}$, and remained in that position until the field reversed direction. For the larger values of $Ma$ in Figure \ref{fig:GdecayPhaseTimeSeries}(c), the form of the sphere response is symmetric under time reversal. However, for $Ma=0.20$, the measured $\theta(t)$ is noticeably asymmetric with respect to time. The origin of the asymmetry is the gravitational torque $\hat{\epsilon}$, which always acts to return the sphere towards $\theta=0$, and this can act with or against the driving magnetic torque at different points in the cycle.

\begin{figure}
	\centering
		\includegraphics[width=0.75\textwidth]{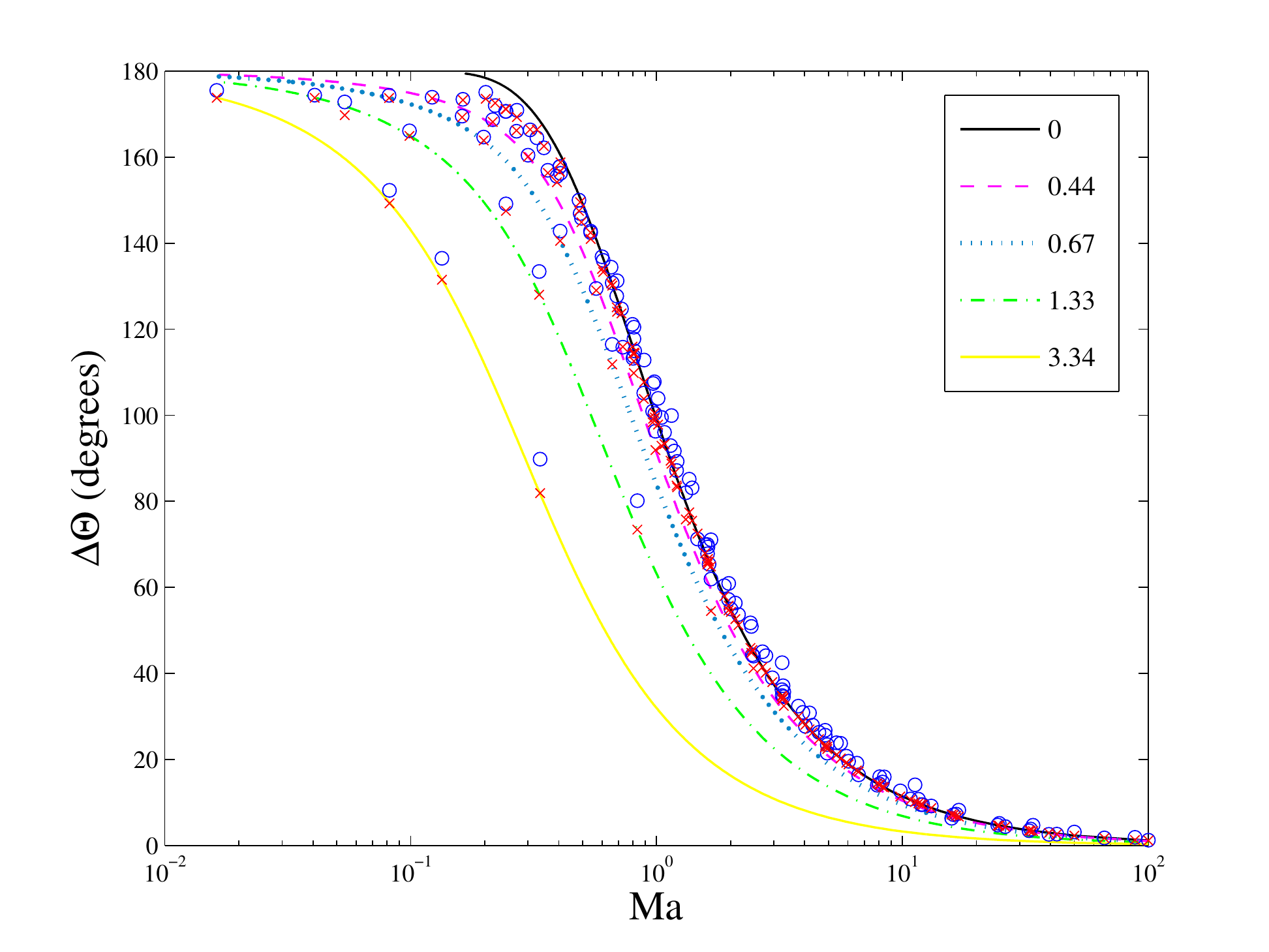}
			\caption{The total angular displacement of the sphere measured as a function of the dimensionless parameter $Ma$ for various $\hat{\epsilon}$. Circles (\textcolor{blue}{o}) represent the experimental data and crosses (\textcolor{red}{x}) represent the numerical solutions evaluated at the corresponding parameters. Error bars on the experimental data points have not been included to aid visual clarity. The analytic solution for the case when the gravitational torque acting on the sphere is zero, $\hat{\epsilon}=0$, is represented by the solid, black line. Numerical results obtained for $\hat{\epsilon}=0.44$, $\hat{\epsilon}=0.67$, $\hat{\epsilon}=1.33$ and $\hat{\epsilon}=3.34$ are represented as indicated in the legend.} 
			\label{fig:AngDispvsFwithbigepsilonhat}
\end{figure}

The total angular displacement of the sphere $\Delta\theta$ was investigated as a function of both the magnetic torque $Ma$, and the gravitational torque $\hat{\epsilon}$, and is shown as a function of $Ma$ for various $\hat{\epsilon}$ in Figure \ref{fig:AngDispvsFwithbigepsilonhat}. The experimental data is shown in Figure \ref{fig:AngDispvsFwithbigepsilonhat} alongside corresponding numerical calculations and the analytic solution to the governing equations of motion obtained for $\hat{\epsilon} = 0$ (see Equation \ref{eq:analyticsol}). Numerical results obtained as a function of $Ma$ for $\hat{\epsilon}=0.44$, $\hat{\epsilon}=0.67$, $\hat{\epsilon}=1.33$ and $\hat{\epsilon}=3.34$ are also included on the figure. Good agreement is found between the experimental and numerical findings when using the full values of $Ma$ and $\hat{\epsilon}$. The experimental data points with small values of $\hat{\epsilon}$ lie close to the analytical prediction equation \ref{eq:analyticsol}, which is the analytical solution in the limit $\hat{\epsilon}=0$.

At fixed $Ma$, an increase in $\hat{\epsilon}$ represents an increase in the gravitational torque acting to return the sphere to the zero-field orientation. This means that greater magnetic torque is required as $\hat{\epsilon}$ increases for the sphere to reach the same alignment with the magnetic field. Increasing $\hat{\epsilon}$ at fixed $Ma$ decreases the amplitude of oscillation and causes the sphere to follow more closely the sinusoidal form of the applied field. An increase in $\hat{\epsilon}$ could be interpreted as an increase in the time-scale of the oscillating field, so that the gravitational torque acts for a longer time period before the field reverses direction. However, changing the frequency of the applied field also results in a change in the Mason number. If the driving frequency is decreased, there is an increase in $\hat{\epsilon}$, and the magnetic field strength must be decreased to maintain the same Mason number.
In order to experimentally investigate the effect of $\hat{\epsilon}$ at fixed $Ma$, and of $Ma$ at fixed $\hat{\epsilon}$, both the amplitude and frequency of the driving magnetic field were varied simultaneously.

Large $\hat{\epsilon}$, for constant $Ma$, significantly reduces the total angular displacement of the sphere, $\Delta \theta$, because gravity acts to return the sphere to the zero-field position. At small $Ma$, $\Delta \theta$ typically is close to $180^\circ$, and is closer to this maximum value when $\hat{\epsilon}$ is small. We can calculate the critical value of $\hat{\epsilon}$ which prevents the maximum angular displacement reaching the maximum value. We initially compute the required $Ma$ which ensured $\Delta \theta$ was within $1\%$ of the maximum response of $180^\circ$ when $\hat{\epsilon}=0$, this was found to be $Ma = 0.21$. We consider the effect of $\hat{\epsilon}$ to be significant when it reduces $\Delta \theta$ by $5\%$. For $Ma=0.21$, this was found to occur at $\hat{\epsilon}=0.396$ through numerical investigation. For $\hat{\epsilon} < 0.396$, the deviation of the angular displacement of the sphere from that predicted by the analytical solution is small and only apparent at angular displacements of close to $180^\circ$, for which gravitational torque prevents the sphere from attaining alignment with the applied field in both directions. However for $\hat{\epsilon}>0.396$, the maximum angular displacement which the sphere can attain is significantly reduced compared to the $\hat{\epsilon}=0$ results.
The experimental and numerical data obtained for $\hat{\epsilon} > 0.396$ shown in Figure \ref{fig:AngDispvsFwithbigepsilonhat} clearly differs from the analytical solution for $\hat{\epsilon}=0$.

\subsection{Flow Field}
\label{subsec:resultsfluid}

Quantitative measurements of the flow were obtained, using the PIV technique outlined in \S\ref{sec:exp}, for a sphere performing small-amplitude torsional oscillations of a sinusoidal form as a result of an applied magnetic field of $0.15$ Hz. The penetration depth of the flow, $\delta \sim (\frac{\nu}{\omega})^{1/2} = 31.3$ mm, was large compared to the radius of the sphere, $\delta \gg a$, which implies that $a^{2}\omega \ll \nu$. Considering that $Re = \omega a^{2}/\nu = 0.064 \ll 1$, we conclude that the motion of the sphere occurred in the low-frequency, low-$Re$ limit \cite{Landau1987}. In the case of low frequencies of oscillation, the fluid velocity varies only slowly with time such that the flow can be regarded as steady at any given instant, and the time-scale associated with the angular acceleration is significantly larger than the viscous time-scale ($a^{2}\nu$) such that the contribution of the acceleration torque to the viscous torque is negligible compared to that of the quasi-steady torque \cite{Lei2006}. 

This is further supported by a consideration of the Reynolds number based instead on the maximum magnitude of fluid velocity, measured to be $v_{max} = 3.85\pm0.17$ mm\,s$^{-1}$, which was found to be $Re_{v} = v_{max}a/\nu \approx 0.03$. The fluid velocity resulting from the torsionally oscillating sphere can therefore be reasonably compared with the analytical solution of the fluid velocity generated by the steady rotation of a sphere in an unbounded, incompressible viscous fluid \cite{Jeffery1915}:
\begin{equation}
	v=\frac{\Omega a^{3}}{r^{2}}\sin\phi
	\label{eq:fluidvelanalyticsol}
\end{equation}
where $\Omega$ is the angular velocity of the sphere, $a$ is the sphere radius, $r$ is the radial distance from the centre of the sphere and $\phi$ is the angle measured from pole-to-pole. Measurements were conducted in the equatorial plane and therefore $\sin\phi=1$.  

\begin{figure}
	\centering
		\includegraphics[width=0.75\textwidth]{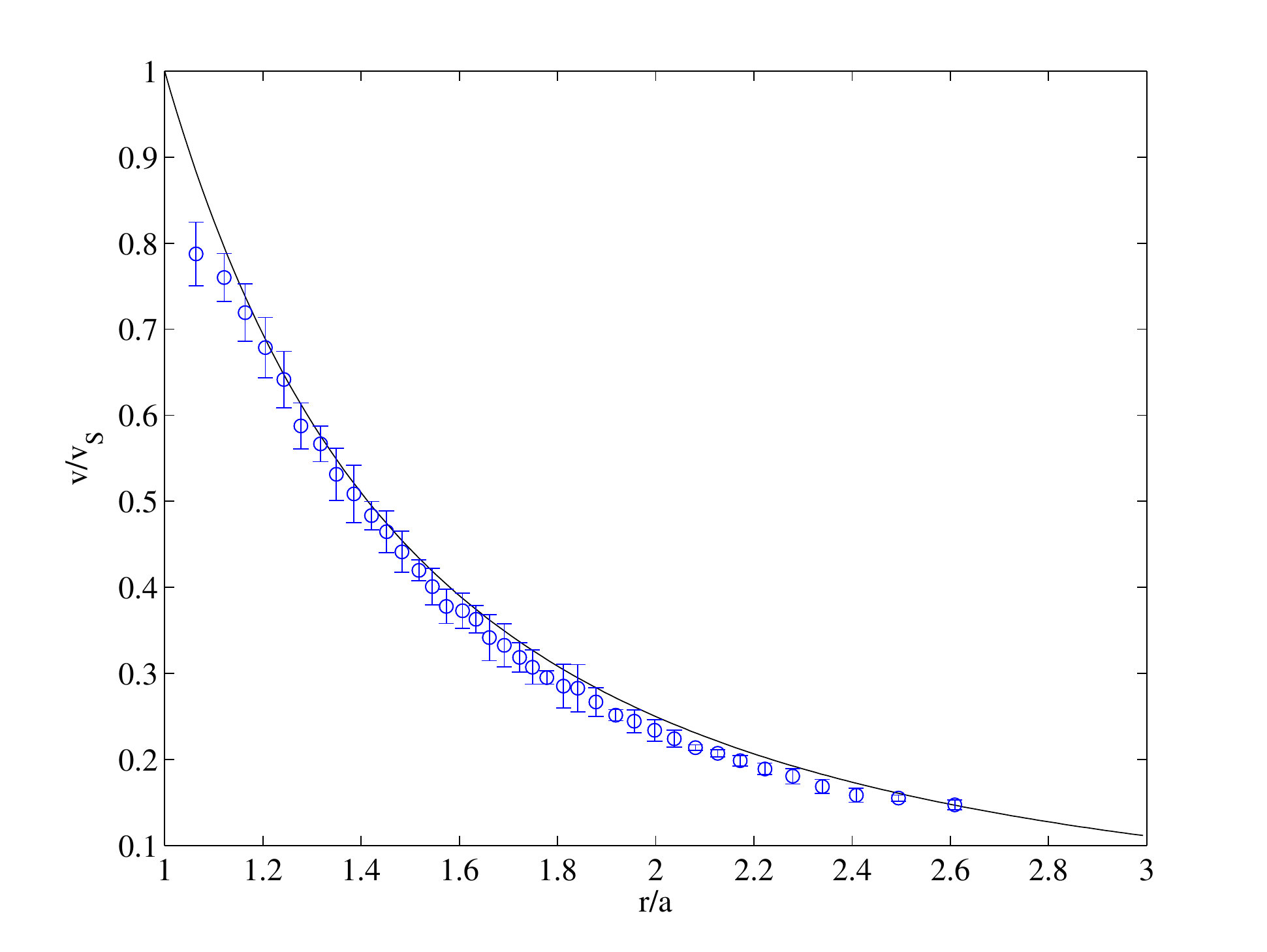}
			\caption{The instantaneous fluid velocity measured as a function of radial distance. The fluid velocity $v$ has been normalised by the surface velocity of the sphere $v_S$, and the radial distance $r$ has been normalised by the radius of the sphere $a$. The black curve is the analytic solution for the fluid velocity due to a sphere rotating with constant angular velocity in an infinite fluid  \cite{Jeffery1915}. The blue data points (\textcolor{blue}{o}) represent the experimental velocity, the standard deviation of which is given by the error bars. Measurements were made for $Ma = 2.01$ and $\hat{\epsilon}=0.22$ for which the sphere performed torsional oscillations of amplitude $\theta = 28^{\circ}$ that approximated a sinusoidal function to within $1\%$.}
		\label{fig:NormalisedVelocityVsRadialDistanceErrorsFitI}
\end{figure}

The magnitude of the fluid velocity was calculated from the experimental vector fields, a typical example of which is shown in Figure \ref{fig:Sch2PIV}(b). The observed symmetry of the flow about the axis of rotation of the sphere enabled spatial averaging around $360^{\circ}$ of the vector field. The empirically measured fluid velocity resulting from the small-amplitude, low-frequency torsional oscillations of a sphere and the fluid velocity calculated from the analytical solution for a sphere undergoing steady rotation, given by equation \ref{eq:fluidvelanalyticsol}, are both shown in Figure \ref{fig:NormalisedVelocityVsRadialDistanceErrorsFitI}. The experimental data points shown in Figure \ref{fig:NormalisedVelocityVsRadialDistanceErrorsFitI} were obtained at $3\pi/10$ in the oscillation cycle yet are typical of instantaneous measurements obtained throughout the period of oscillation. 

Good quantitative agreement is found between the experimental and analytical fluid velocity as a function of radial distance. This validates the Stokes flow assumption used to develop the model described in \S\ref{sec:theory} which, in deriving the viscous torque acting on the sphere, assumes that $Re=0$ and so corresponds to the velocity field given by equation \ref{eq:fluidvelanalyticsol}. The consistent underestimation of the measured fluid velocity, of $\approx 5\%$, is attributed to the finite width of the laser-sheet illumination. The significant deviation of the experimental data from the analytic solution close to the sphere results from specular reflections by the surface of the sphere which, prior to diffusing, illuminate tracer particles that are not on the equatorial plane. Whereas, the marginal reduction in agreement at a radial distance $\gtrsim 2a$ is attributed to the influence of the boundaries of the tank containing the viscous fluid which were $\approx 8a$ from the centre of the sphere.

\section{Summary}

A near neutrally buoyant sphere with a magnetic-dipole axis and a non-uniform mass distribution was submerged in a viscous fluid and subjected to an periodically-oscillating magnetic field. The sphere performed torsional oscillations about the zero-field position. A simple balance of the magnetic, gravitational and viscous torques acting on the sphere enabled identification of two non-dimensional parameters which determined the behaviour of the sphere: The Mason number, $Ma$, which quantified the ratio of viscous to magnetic torque; and $\hat{\epsilon}$, which quantified the ratio of the gravitational to viscous torque. The interaction of an applied magnetic field with the magnetic dipole within the sphere resulted in a magnetic torque which acted to align the magnetic axis of the sphere with the applied field. Viscous drag, arising from the no-slip condition on the surface of the sphere, resisted the motion of the sphere. The effects of gravity on the non-uniform mass distribution within the sphere acted to return the sphere to the zero-field position. The zero-field position of the magnetic-dipole axis of the sphere was approximately at $\theta=0$ because of careful embedding of the weights used to achieve neutral buoyancy of the sphere. The response of the sphere to an applied field was investigated experimentally and numerically and good agreement was found between the experimental data and the numerical results. 

Characterising the response of the sphere to the applied magnetic field enabled determination of the parameter regime for which the sphere performs small-amplitude torsional oscillations of a sinusoidal form. Knowledge of the resulting angular displacement of the sphere provides the basis for oscillating sphere viscometers for medical, biological and micro-fluidic applications. The advantage of a viscometer based on rotational motion, compared to conventional methods based on translational motion, is that a smaller volume of fluid is probed. This enables the measurement of the rheological properties of micro- to pico-litre volumes of viscous fluids \cite{Besseris1999} and \textit{in vivo} testing \cite{Parkin2007}. Furthermore, non-mechanical control is essential in such viscometers, as the friction between the rotational axis of the sphere and the mechanical support often restricts the lower limit of viscosity measurement \cite{Sakai2010}.

The flow generated by the motion of the sphere was studied in the equatorial plane and found to move in arcs along a circular trajectory around the sphere, the direction of which was determined by the rotation of the sphere. The fluid velocity was found to have a functional dependence on the radial distance from the sphere, decreasing with increasing radial distance. For low frequency oscillations ($a^{2}\omega \ll \nu$) of small amplitude at low-Reynolds number ($v_{max} a/\nu \ll 1$), such that the flow can be regarded as steady at any given instant \cite{Landau1987}, good agreement was found between experimental measurements of the fluid velocity around the sphere and the analytical solution of the functional dependence of the fluid velocity on radial distance from a sphere performing steady rotational oscillations in an infinite, viscous fluid. This confirms that the set up is an experimental realisation of a Stokes flow in which the oscillations of particles can be actuated via a non-contact method. The experimental apparatus thus provides a controlled environment in which other phenomena, such the hydrodynamic interaction between oscillating spheres, may be explored. 

\section*{Acknowledgment}

FB would like to acknowledge the technical assistance of P. Tipler and to thank Professor K. Novoselov and Professor A. Murray for the loan of equipment.  The completion of the experimental work was supported by an EPSRC studentship (EP/P505631/1), and the writing of the manuscript by a David Crighton Fellowship.


\end{document}